\pdfoutput=1
\documentclass[preprints,article,accept,moreauthors,pdftex,10pt,a4paper]{mdpi} 

\preto{\abstractkeywords}{\nolinenumbers}
\firstpage{1} 

\history{February 2019}

\usepackage{multirow} 
\usepackage{xspace}
\usepackage{overpic}
\usepackage{gensymb}
\usepackage[caption=false]{subfig}
\usepackage{bookmark}
\usepackage{amssymb}

\Title{Slow Waves Analysis Pipeline for extracting the Features of the Bi-Modality from the Cerebral Cortex of Anesthetized Mice}

\Author{Giulia De Bonis $^{1}$, Miguel Dasilva $^{2}$, Antonio Pazienti $^{3}$
		Maria V. Sanchez-Vives $^{2}$, Maurizio Mattia $^{3}$ and Pier Stanislao Paolucci $^{1}$}
\AuthorNames{Giulia De Bonis, Miguel Dasilva, Antonio Pazienti, Maria V. Sanchez-Vives, Maurizio Mattia and Pier Stanislao Paolucci}

\address{%
$^{1}$ \quad Istituto Nazionale di Fisica Nucleare (INFN), Sezione di Roma, Rome, Italy\\
$^{2}$ \quad Institut d'Investigacions Biom\`ediques August Pi i Sunyer (IDIBAPS), Barcelona, Spain\\
$^{3}$ \quad Istituto Superiore di Sanit\`a (ISS), Rome, Italy}

\corres{Correspondence: giulia.debonis@roma1.infn.it}

\abstract{Cortical slow oscillations ($\lesssim 1\ Hz$) are an emergent property of the cortical network that integrate connectivity and physiological features. This rhythm, highly revealing of the characteristics of the underlying dynamics, is a hallmark of low complexity brain states like sleep, and represents a default activity pattern.
Here, we present a methodological approach for quantifying the spatial and temporal properties of this emergent activity. We improved and enriched a robust analysis procedure that has already been successfully applied to both \textit{in vitro} and \textit{in vivo} data acquisitions. 
We tested the new tools of the methodology by analyzing the electrocorticography (ECoG) traces recorded from a custom 32-channel multi-electrode array in wild-type isoflurane-anesthetized mice.
The enhanced analysis pipeline, named SWAP (Slow Waves Analysis Pipeline), detects Up and Down states, enables the characterization of the spatial dependency of their statistical properties, and supports the comparison of different subjects. The SWAP is implemented in a data-independent way, allowing its application to other data sets (acquired from different subjects, or with different recording tools), as well as to the outcome of numerical simulations.
By using SWAP, we report statistically significant differences in the observed slow oscillations (SO) across cortical areas and cortical sites. Computing cortical maps by interpolating the features of SO acquired at the electrode positions, we give evidence of gradients at the global scale along an oblique axis directed from fronto-lateral towards occipito-medial regions, further highlighting some heterogeneity within cortical areas.
The results obtained using SWAP will be essential for producing data-driven brain simulations. A spatial characterization of slow oscillations will also trigger a discussion on the role of, and the interplay between, the different regions in the cortex, improving our understanding of the mechanisms of generation and propagation of delta rhythms and, more generally, of cortical properties.}
\begin{document}
\section{Introduction}
\label{sec:Intro}
In a brain manifesting slow-wave activity (SWA), expressed in the cerebral cortex under NREM sleep and deep anesthesia \citep{Steriade1993:SlowOscillations}, the spiking activity -- both single and multi-unit activity (SUA and MUA respectively) -- appears as a regular sequence of Up (high-rate) and Down (almost quiescent) states. 
In the past 10 years, an accurate procedure for the analysis of electro-neurophysiological data has been developed, aimed at extracting the MUA from raw recordings, identifying alternating Up and Down states associated with SWA and investigating the features of such rhythmic spatio-temporal patterns of activity propagating along the cortex. 
The analysis pipeline, implemented in MATLAB\footnote{MATLAB\textsuperscript\textregistered, The MathWorks, Inc., Natick, Massachusetts, United States.}, has been properly refined over time, and applied to several experimental data sets, acquired with different setups both \textit{in vitro} and \textit{in vivo} from rodents, ferrets, and monkeys \citep{Ruiz-Mejias2011:AnesthetizedMouse,Mattia2013:MotorPlanningPremotorCortex,Capone2017:SlowWavesCorticalSlices,Ruiz-Mejias2016:Dyrk1A}.
The software procedure has been extensively revised and improved, including some new features implemented in Python\footnote{Python Software Foundation. Python Language Reference, version 2.7.5. Available at \url{http://www.python.org}.}, to be used with a new set of data collected \textit{in vivo} using a 32-electrode array from the cerebral cortex of 11 deeply isoflurane-anesthetized wild-type mice. 
The data used in this study was obtained in accordance with the Spanish regulatory laws (BOE-A-2013-6271) which comply with the European Union guidelines on protection of vertebrates used for experimentation (Directive 2010/63/EU of the European Parliament and the Council of 22 September 2010), and the protocol was approved by the Animal Ethics Committee of the University of Barcelona.

\begin{figure}[h!]
    \centering
    \begin{overpic}[width=0.48\textwidth,trim={0cm -3cm 0cm 0cm},clip]{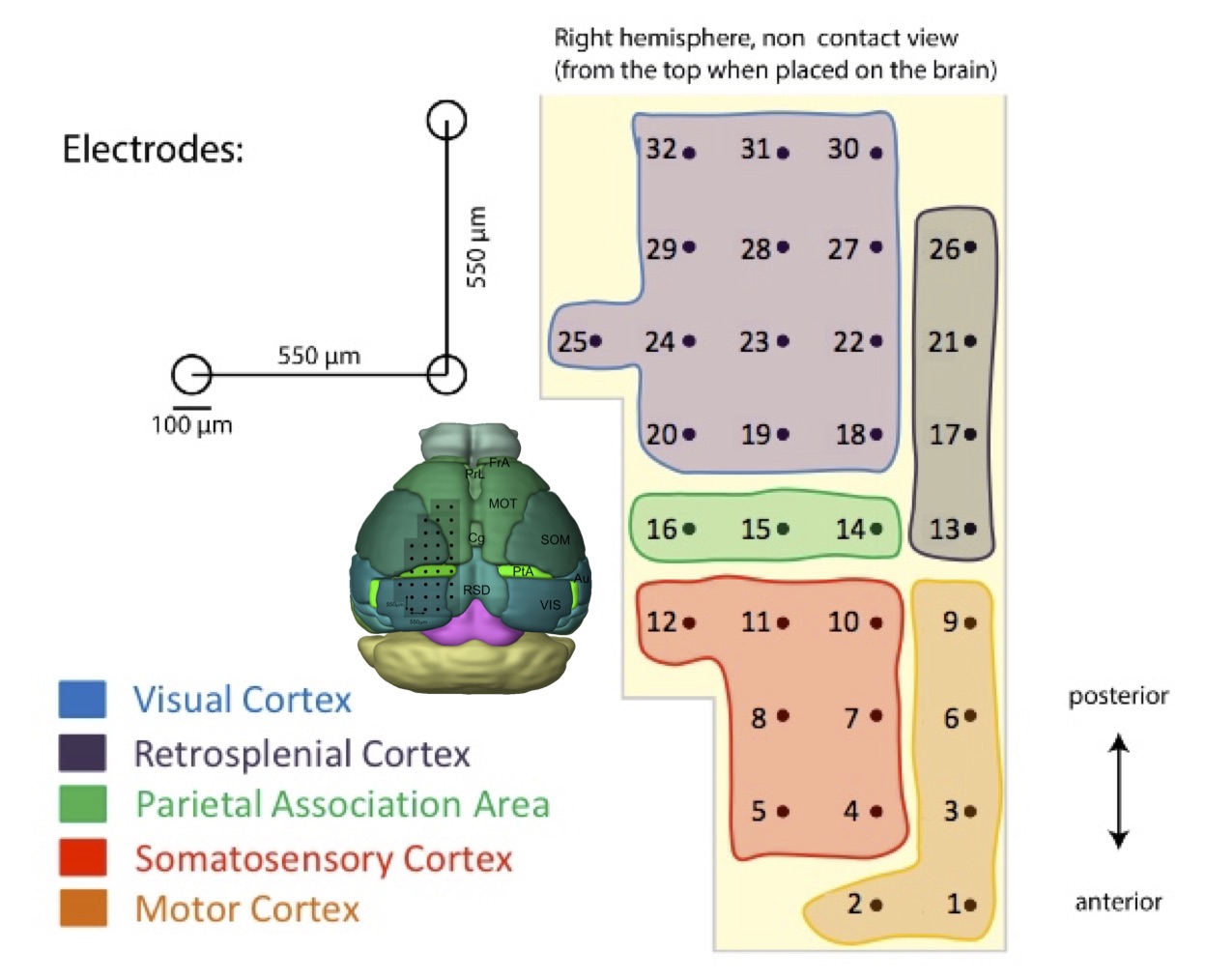}
        \put(60,5){\fontsize{6pt}{8pt}\selectfont{lateral $\longleftrightarrow$ medial}}
    \end{overpic}
    \caption{Representation of the multi-electrode array (MEA) used for the data acquisition \cite{Pazzini2018:UltraCompactSystem}. On the top left, a scheme that indicates the scale of the experiment; the reported dimensions have been adopted to define the reference frame in the SWAP. On the bottom left, the color legend that identifies the cortical areas on the array surface. On the right, the grid of electrodes, with the numbering introduced in the SWAP. In the center, an illustration of the MEA positioned on the mouse cortex; the inspected surface is of the order of $10\ mm^2$.}
    \label{fig:ElectrodeArray}
\end{figure}

The multi-electrode array (MEA) employed for acquiring the data is described in Figure \ref{fig:ElectrodeArray} and covers several cortical areas ranging from sensory (V1, S1), motor (M1) and association (PtA) cortices \cite{Pazzini2018:UltraCompactSystem}. The unfiltered field potential (UFP) is sampled from each electrode at a frequency of $5~kHz$ and each acquisition session lasts at least $300~s$ (see Table \ref{tab:DAQsession-details}), thus ensuring a fine inspection of the signal in both space and time.
The 11 recording sessions are each from a different animal, \textit{i.e.} 11 independent experiments that collectively represent an extensive data sample covering a wide range of biological and unavoidable physiological variability. 

\begin{table}[h!]
\centering
\begin{tabular}{|p{1.25cm}|p{1.6cm}|p{2.2cm}|p{8.75cm}|}\hline 
\vspace{2pt} Data File & \vspace{2pt} Hemisphere & Duration of the DAQ Session $[s]$ & \vspace{2pt} \hspace{3.5cm} Notes \\[6pt]\hline
\vspace{6pt}01 & \vspace{6pt}R & \vspace{6pt}304.102 & \begin{small}\begin{itemize}\item Excluded: ch. 8 (SD-outlier) \end{itemize} \end{small} \\ \hline
\vspace{15pt}02 & \vspace{15pt}R & \vspace{15pt}300.912 & \begin{small}\begin{itemize}\item Weak Bimodality: ch. 1, 2, 3, 4, 7, 9, 10, 11, 12, 14, 15, 16, 17, 18, 19, 20, 22, 23, 24, 25, 26, 27, 28, 29 \item Excluded: ch. 30, 31, 32 (failure in the DAQ) \end{itemize} \end{small}\\ \hline
\vspace{12pt}03 & \vspace{12pt}L & \vspace{12pt}427.223 & \begin{small}\begin{itemize}\item Discontinuity in ch. 2, 4, 8, 12, 15, 19, 20, 23, 28, 31, 32 \item Excluded: ch. 3, 13 (SD-outlier) \end{itemize} \end{small}\\\hline
\vspace{1pt}07 &\vspace{1pt} L & \vspace{1pt}351.701 & \vspace{3pt}-- \\[6pt]\hline
\vspace{15pt}09 & \vspace{15pt}L & \vspace{15pt}321.942 & \begin{small}\begin{itemize} \item Discontinuity in ch. 23 \item Weak Bimodality: ch. 2, 3, 17, 20 \item Excluded: ch. 3, 9 (SD-outlier)\end{itemize} \end{small} \\\hline
\vspace{12pt}10 & \vspace{12pt}L & \vspace{12pt}309.19 & \begin{small}\begin{itemize} \item Discontinuity in ch. 12 \item Excluded: ch. 12 (failure in the DAQ) \end{itemize} \end{small} \\\hline
\vspace{1pt}14 & \vspace{1pt}R & \vspace{1pt}313.714 & \vspace{3pt}-- \\[6pt]\hline
\vspace{6pt}15 & \vspace{6pt}R & \vspace{6pt}329.605 & \begin{small}\begin{itemize} \item Excluded: ch. 25 (SD-outlier) \end{itemize} \end{small} \\\hline
\vspace{12pt}16 & \vspace{12pt}R & \vspace{12pt}312.157 & \begin{small}\begin{itemize} \item Discontinuity in ch. 14 \item Excluded: ch. 4 (SD-outlier); 31 (failure in the DAQ)  \end{itemize} \end{small} \\\hline
\vspace{1pt}17 &\vspace{1pt} R & \vspace{1pt}305.967 &\vspace{3pt} -- \\[6pt]\hline
\vspace{6pt}20 & \vspace{6pt}R & \vspace{6pt}319.942 & \begin{small}\begin{itemize} \item Excluded: ch. 1 (negative asymmetry); 12, 25 (SD-outlier)  \end{itemize} \end{small} \\\hline
\end{tabular}
\caption{Summary of the data acquisition (DAQ) sessions. The name of the data file reflects the date of acquisition. $R$ for right hemisphere, $L$ for left hemisphere. In general, discontinuities are not critical, since the failure corresponds to an interruption of the DAQ for a limited time interval (usually, a couple of discontinuities per channel, lasting from a few hundreds of millisecond to a few seconds); therefore, discontinuities in the DAQ are managed in the SWAP by identifying, for each problematic channel, the time interval at which the data acquisition fails, and removing it from the time sequence of the signal. By contrast, excluded channels are those for which the signal presents several irregularities, usually resulting in a number of identified upward transitions well below the median computed over the full channel set (tagged as ``failure in the DAQ''); in addition, SD-outlier channels are also excluded; the tag ``negative asymmetry'' corresponds to the case of having a strong negative skew.}
\label{tab:DAQsession-details}
\end{table}

The accurate time-and-space sampling together with the richness of the experimental data have driven the development of new analytical tools that aim to characterize the differences between cortical areas when expressing SWA. In addition, given the unavoidable and physiological variability of the data set, particular attention has been also given to the best strategies to adopt in order to perform a thorough comparison of recordings obtained from the 11 different subjects. The guiding principle when combining data was to keep and use the largest amount of signal, avoiding arbitrary removal of noisy channels or the creation of a subset of ``golden'' cases.
Descriptive statements are accompanied by the assessment of statistical significance of the results, taking into account the multiplicity of the hypotheses under testing; the obtained claims can give hints as to the mechanisms underlying SWA in mammals.

Furthermore, the outcome of the data analysis can be employed to feed the input of a dedicated spiking neural network simulation (as \citep{Paolucci2013:DPSNN,Pastorelli2015:DPSNN,Pastorelli2018:LateralConnectivity-PDP}), in a data-driven approach of \textit{in silico} studies of the brain, aimed at computing a less stereotypical and a more accurate reproduction of cerebral rhythms. The analysis pipeline itself, hereafter named Slow Waves Analysis Pipeline, or SWAP, can be adopted to study the output of the simulation, and to define a set of benchmark observables for confronting models and experiments, with the goal of having a reliable and flexible set of tools available for the characterization of the slow-wave signal in a wide set of cases. 

The material in this paper is structured as follows. 
Section \ref{sec:SWAP-overview} offers an overview of the SWAP analysis pipeline, with a collection of results obtained from test-bench data used for illustrating the steps of the procedure; the focus is on the methods implemented for the identification of the Up/Down state alternation in the ECoG traces, on the techniques for ``stacking'' and comparing different subjects, and on the statistical treatment of data with hypothesis testing.
Section \ref{sec:discussion} is dedicated to the discussion of methods and results, with concluding remarks and suggestions for future research.

\section{The Slow Waves Analysis Pipeline (SWAP)} 
\label{sec:SWAP-overview}

The study of SWA expressed by the cortex can be tackled starting from a description of the bimodality (\textit{i.e.} the alternation of Up and Down states), by defining a comprehensive set of observables suited to illustrating the phenomenon of SO.
Once this local information is acquired, the second step (not illustrated here) is the characterization of the activity propagation across the brain surface as a wave with delta rhythm. 
Focusing on the features of Up and Down states, differences between cortical sites can be emphasized.

The analysis procedure consists of two phases. First, each data file is examined separately (\textit{Single Experiment}), yielding a detailed inspection of the single subject. The results from the different experiments are then combined (inter-session data) to produce \textit{Summary Results} . Finally, conclusive claims are given with the assessment of statistical significance of the results, taking into account the numerousness of the sample and the multiplicity of the hypotheses under testing. 

\begin{figure}[h!]
\begin{center}
\includegraphics[scale=0.305]{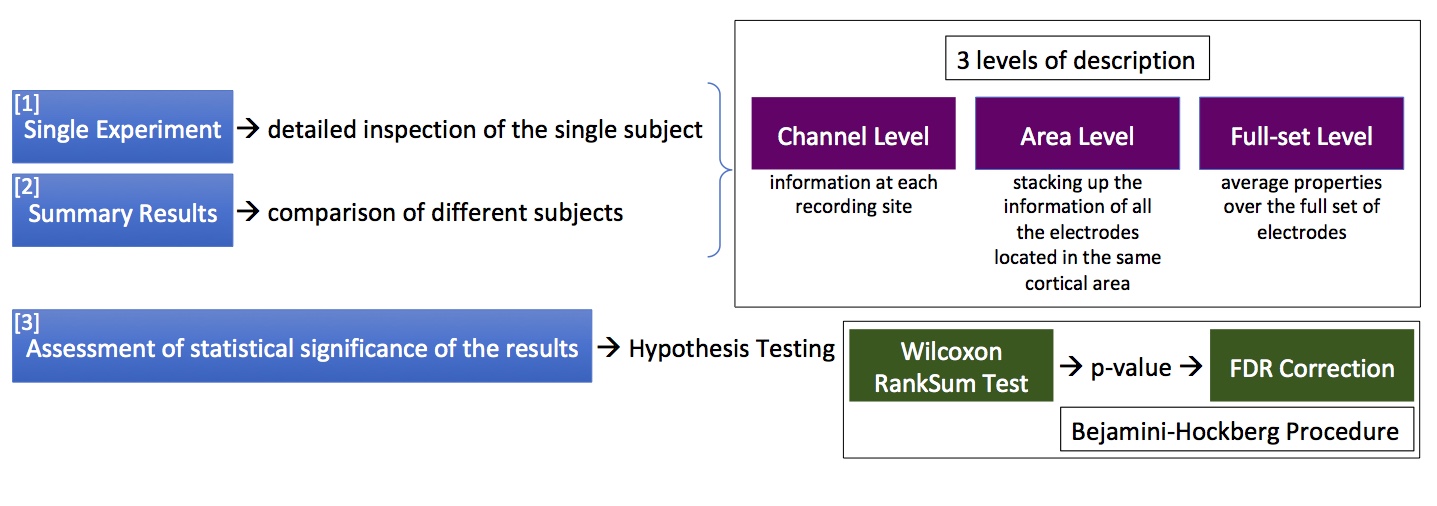}
\end{center}
\caption{The logic of the SWAP when facing the issue of investigating the features of Up and Down states, in terms of phases of action (the sequence of 3 actions listed on the left) and levels of description (in the upper box). The lower box contains information on the statistical treatment presently implemented.}
\label{fig:SWAP_issue1_phases-and-levels}
\end{figure}

Three different levels of description can be enabled: (I) the channel level, providing information at each recording site; (II) the area level, stacking up the information of all the electrodes located in the same cortical area; and (III) the full-set level, computing average properties and summary statements for the entire cortex portion under study. 
The levels of description can be applied at the single experiment, or at the inter-session data; outcomes obtained at the different levels of description are complementary and not strictly separated, and can be superimposed on graphical representations. Figure \ref{fig:SWAP_issue1_phases-and-levels} details, in terms of phases of action and levels of description, the logic of the SWAP when facing the issue of investigating the features of Up and Down states. 

\subsection{The channel-level description of the single experiment: from the raw data to the estimates of MUA and of transition times}
\label{subsec:channel-level_SingleExp}
\begin{figure}[h!]
\begin{center}
\includegraphics[scale=0.305]{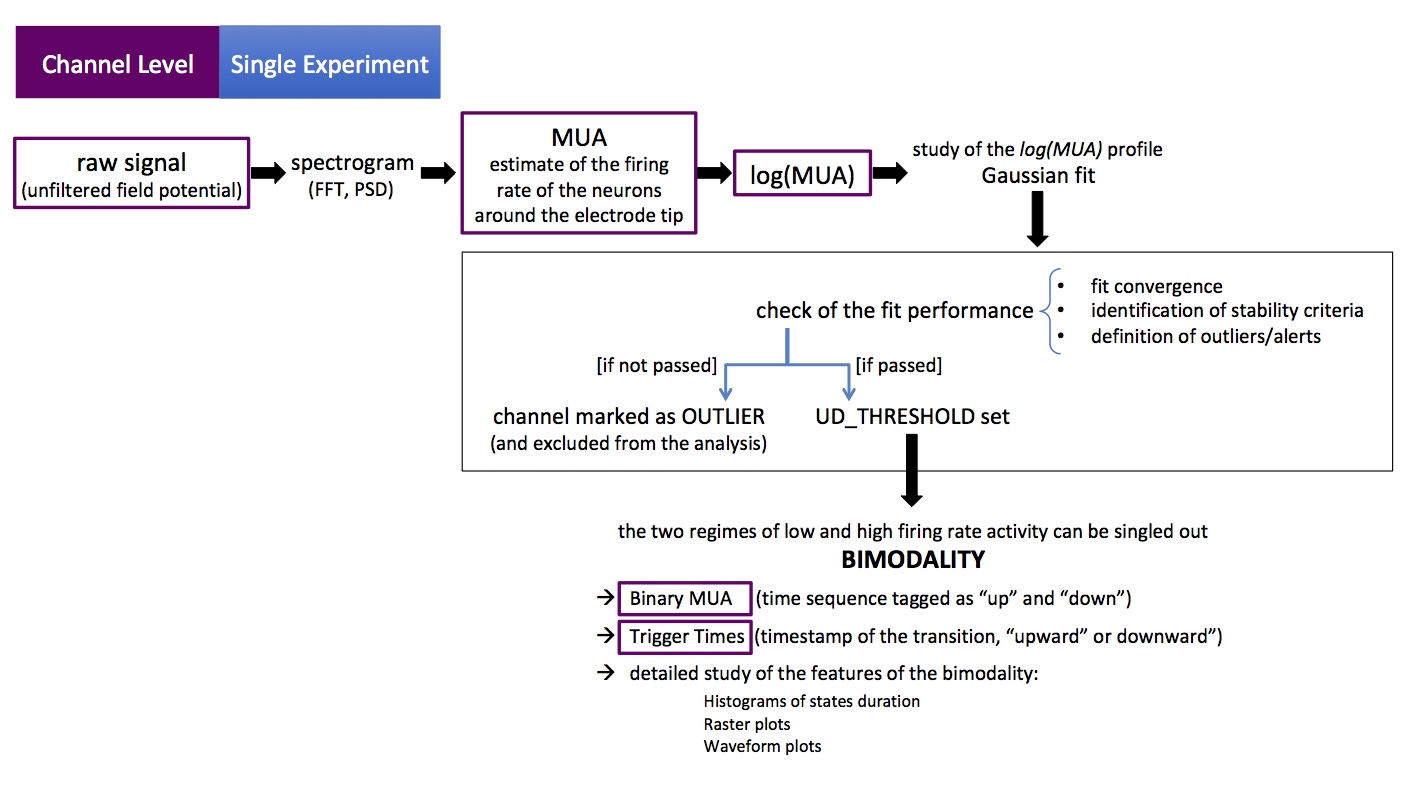}
\end{center}
\caption{The channel-level description of the single experiment. Purple boxes point out the initial data format and the intermediate outcomes of the workflow (as successive elaborations of the raw data). The black box encompasses the key steps of the process. The procedure is coded in MATLAB\textsuperscript\textregistered, the sequence of actions is illustrated in Figure \ref{fig:ChannelLevel-SingleExp_Illustration}. The figure illustrates the main loop (as discussed in Section \ref{subsec:channel-level_SingleExp}), at the end of which further checks are carried out on average and median values (full-set level description), in order to identify further anomalies or outliers.}
\label{fig:ChannelLevel-SingleExperiment}
\end{figure}

The channel-level description of the single experiment is obtained with a set of scripts, coded in MATLAB\textsuperscript\textregistered, carrying out a loop over the electrodes in the array to extract the MUA from recordings of the raw signal (UFP). For each channel, the Power Spectral Density (PSD) of the signal is computed and the MUA is used as an estimate of the firing rate of the neurons around the electrode tip \cite{Ruiz-Mejias2011:AnesthetizedMouse,Mattia2010:MotorDecisionTask}. The logarithm of the MUA is evaluated, and the shape of the $log(MUA)$ distribution can be described as a \textit{peak}, at low-MUA values corresponding to Down states, and a \textit{tail}, at high-MUA values corresponding to the Up states. The $log(MUA)$ peak is fitted with a Gaussian function, and the parameters of the fit are used to single out Down-state periods from Up-state periods in the MUA time series. Once the MUA time series is tagged as ``Up'' and ``Down'' (binary MUA), a detailed study of the features of such states and of the transitions among them -- upward (Down-to-Up) and downward (Up-to-Down) -- can be achieved.
The workflow is illustrated in Figure \ref{fig:ChannelLevel-SingleExperiment}.

\begin{figure}[h!]
    \centering
    \begin{overpic}[width=1\textwidth]{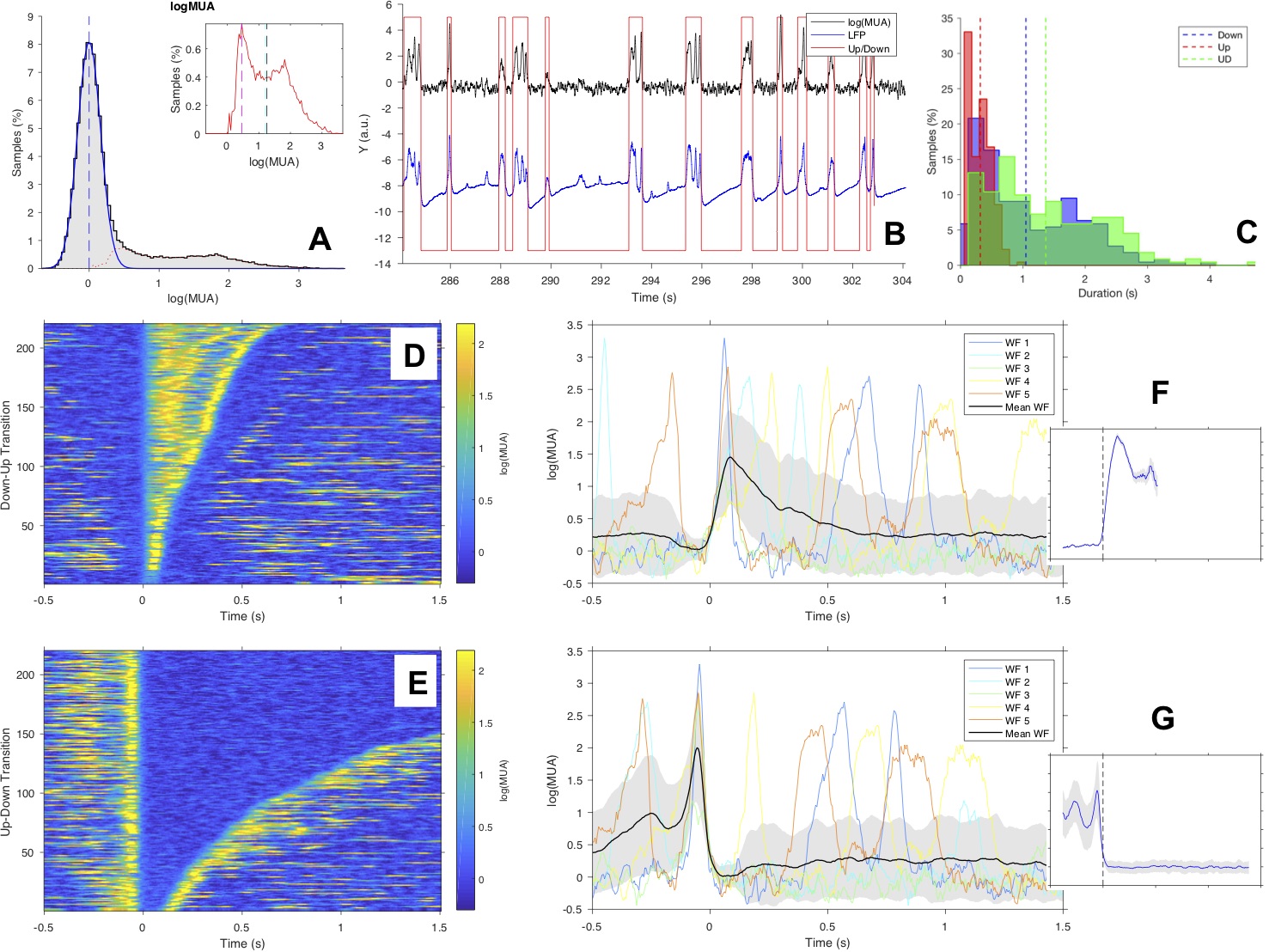}
        \put(87.9,3.15){\fcolorbox{white}{white}{\fontsize{5pt}{7pt}\selectfont{Time (s)}}}
        \put(87.9,28.85){\fcolorbox{white}{white}{\fontsize{5pt}{7pt}\selectfont{Time (s)}}}    
    \end{overpic}
    \caption{\textbf{(A)} Distribution of $log(MUA)$ as normalized histogram of values shifted at the position of the first peak, in order to set at level $0$ the average firing rate that corresponds to a Down state, thus associated with a null firing rate.
    The blue curve is the Gaussian fit of the first peak, the $\mu$ position is marked by the blue vertical dashed line. The red dotted line is the ``tail'', obtained after subtracting the Gaussian fit from the total distribution (see inset; vertical dashed lines: mode in magenta, median in cyan, mean in black).
    In this example channel (Channel 1, file 01), the tail distribution is far from being Gaussian;
    the area of the tail is $\sim27\%$, therefore expressing a clear bimodality, coherently with the value of the skewness of the distribution, $\gamma=1.97$.  \textbf{(B)} $log(MUA)$ (in black), \textit{BinaryMUA} (in red) and raw signal (UFP, in blue). \textbf{(C)} Histograms of the duration of Up states (red), of Down states (blue) and of the full cycles (UD cycle, in green); dashed lines, mean values. \textbf{(D,E)} Raster plot of upward (Down-to-Up) and downward (Up-to-Down) transitions, sorted by state duration. \textbf{(F,G)} Waveforms (WFs) of upward and downward transitions, respectively. For each plot, the first 5 transitions are shown, centered around the transition time (\textit{Trigger Time}); the average transition, computed considering the full set of $n$ waveforms, is superimposed (black); the shaded area identifies the profile of the waveform at $\pm 1$ standard deviation (SD) off the mean. The inset shows the average transition front; the shaded area identifies the profile of the waveform at $\pm 1$ standard error (SEM) of the mean (\textit{i.e.} smaller error in the proximity of the trigger time and larger far from it, since moving away from the transition front the fluctuations increase and the number of events contributing to the mean decreases).}
    \label{fig:ChannelLevel-SingleExp_Illustration}
\end{figure}

In more detail, the initialization phase is carried out with the steering file \textit{setParamsAndOptions}, which takes into account the specificity of the data acquisition (DAQ) sessions, accommodating the frame of reference (the positions of the electrodes in the array) and loading information about the recordings.
Some of the settings concern the capability of the algorithm to identify the state transitions, and are fine-tuned following a heuristic approach. The steering file informs what to check in order to perform the analysis pipeline on the given input data.
Once the initialization phase is completed, the main loop starts and the flow in the pipeline is as follows (see Figure \ref{fig:ChannelLevel-SingleExp_Illustration}):

\begin{enumerate}
    \item Read the analog data (raw signal). For the ECoG data used as a benchmark in this study, input-related actions (open/close the input file and read the input data) are performed by the SONLibrary \cite{Lidierth2009:SON}, whose functions provide the access to the stored neurophysiological data. 
    \vspace{0.15cm}
    \item Analyze the raw signal in the frequency domain. A moving window of samples is defined for the calculation of the spectrogram, \textit{i.e.} the spectral content of the raw signal as a function of time; the extent of the moving widow, together with the sampling frequency, determines the frequency band to be examined. Since the frequency band of interest for the estimate of the MUA is ($200-1500$)~$Hz$\footnote{The experience suggests that frequencies in the raw signal outside this range cannot be associated with the electro-physiological signal of the MUA; in particular, the highest frequency components reflect the presence of electrical noise introduced by the acquisition system.}, a moving window of $5$~$ms$ is adopted\footnote{For the benchmark data acquired at $5~kHz$, the time window contains 25 samples.}. FFT (Fast Fourier Transform) and PSD (Power Spectral Density) are computed using  MATLAB\textsuperscript\textregistered functions. Then, for each frequency in the band, the median of the PSD is computed, considering the full set of spectrograms, \textit{i.e.} the collection of moving windows (time steps) that constitute the entire acquisition session. The obtained vector of medians, one value for each frequency, is used as a baseline to normalize the PSD. 
    \vspace{0.15cm}
    \item Evaluate the MUA for each time step as the mean amplitude of the normalized PSD. The MUA is intended as an estimate of the collective firing rate $r(t)$ of neurons located at the electrode position. The natural logarithm of the MUA is computed, since logarithmic mapping is adopted to emphasize the bimodality of the distribution; because of the normalization to the median of the PSD, negative values and positive values of $log(MUA)$ identify a spectral content smaller or larger than the median, respectively \cite{Ruiz-Mejias2011:AnesthetizedMouse,Mattia2010:MotorDecisionTask}. 
    \vspace{0.15cm}
    \item Fit the distribution of $log(MUA)$, isolating with a Gaussian function the peak corresponding to the (dominant) regime of low-rate states (Figure \ref{fig:ChannelLevel-SingleExp_Illustration}.A). A simplified description of the bimodality of the SWA in the cortex would require a bimodal fit, and initially the sum of two Gaussian profiles was adopted to describe the shape of the distribution. Conversely, what was discovered after a thorough inspection of a huge number of channels from several different animals (both physiological and pathological subjects at different anesthesia levels) is that, while the Down-state peak is highly stable despite the large variability in the subjects, the content of the $log(MUA)$ histogram corresponding to the high-rate regime (obtained from the total distribution after subtracting the Down-state peak) expresses over a large span of ``shapes'', and rather than as a ``second peak'' (with a definite Gaussian profile) it can be generically appointed as a ``tail'' (Figure \ref{fig:ChannelLevel-SingleExp_Illustration}.A, inset).
    \vspace{0.15cm}
    \item Set the \textit{UD\_THRESHOLD}, \textit{i.e.} the level of $log(MUA)$ that defines the separation of the two regimes of the bimodality, Up and Down (UD). 
    This is a crucial step in the pipeline, and several options can be adopted to find the optimal criterion, which can depend on the specificity of the data set and on the scope of the analysis. In general, the choice takes into account general settings of the DAQ, the $log(MUA)$ distribution, and the results of the Gaussian fit on the Down-state peak (of the given channel, or considering average properties of the recording session). 
    At the channel level in the main loop, checks are introduced to monitor the convergence of the Gaussian fit, the content of the histogram of $log(MUA)$ values, and the shape of the distribution (Figure \ref{fig:ChannelLevel-SingleExp_Illustration}.A); alerts are activated to signalize: \textit{Weak Bimodality} (if the area of the tail is smaller than a given threshold -- currently set at 10\% of the total); \textit{Positive Skewness} (if gamma -- the coefficient of skewness -- of the tail is above $1$, $\gamma>1$); \textit{Negative Skewness} (if $\gamma<-1$); \textit{Right Peak} (if the peak, \textit{i.e.} the dominant component of $log(MUA)$, is centered at large values, on the right segment of the range, with a tail on the left); \textit{Large Threshold} (if the selected threshold is larger than the mean of the tail); \textit{Few Transitions} (if the number of transitions obtained with the selected threshold is below $3$, the minimum requisite to isolate at least an Up state and a Down state). Some of these alerts may help in defining the threshold level, others provide an indication of ``problematic'' channels to be inspected (that may or may not be tagged as outliers and excluded from further processing at the end of the main loop). Finally, some conditions are ``blocking'', in the sense they require a break in the workflow, as is the case with noisy acquisition channels or spoiled recordings. If no blocking conditions are encountered, meaning that a threshold can be set and Down states and Up states are distinguishable, the following operations are performed:
	\begin{enumerate}
		\item Convert the MUA into a binary sequence (\textit{BinaryMUA}) that is $1$ or $0$ depending on the value of $log(MUA)$ above or below the threshold, resulting in the MUA time sequence being tagged as ``Up'' or ``Down'' (Figure \ref{fig:ChannelLevel-SingleExp_Illustration}.B).   
		\vspace{0.15cm}
        \item Label the transitions as upward or downward, and assign the time of transition (\textit{Trigger Time}) with a cubic interpolation of the waveform to locate the time at which the level of MUA crosses the threshold that separates the two regimes of low and high firing rate activity.
        \vspace{0.15cm}  
        \item Study the features of states and transitions, to check the robustness of the algorithm and the stability of the outcome for the different channels. In particular: 
        \begin{enumerate}
            \item Histograms of the duration of Down states $d_{DOWN}$, Up states $d_{UP}$, UD-cycles $d_{UD}$. The study of these observables is one of the focuses of the analysis; at the channel-level description of the SWAP, the distributions are superimposed for comparison (Figure \ref{fig:ChannelLevel-SingleExp_Illustration}.C). 
            \vspace{0.15cm}
            \item Raster plots of states and transitions; each transition (\textit{Trigger Event}) is centered at the trigger time; events can be sorted by their time of occurrence, or by their duration (Figure \ref{fig:ChannelLevel-SingleExp_Illustration}.D-E). 
            \vspace{0.15cm}
            \item Waveform plots, for comparing states and transitions, and plots of the average waveform, obtained considering the full set of transitions. A refined algorithm isolates the transition front, to better investigate the transition dynamics (Figure \ref{fig:ChannelLevel-SingleExp_Illustration}.F-G). 
        \end{enumerate}
	\end{enumerate}
\end{enumerate}

As anticipated above, once the main loop execution has yielded a full description at the channel level, a key parameter to be monitored for the validation of the procedure is the stability of the conditions used for the identification of the two states (low-MUA and high-MUA). More precisely, since a requirement for the separability of Up and Down states is the successful fit of the Down-state peak of the $log(MUA)$ with a Gaussian function (Figure \ref{fig:ChannelLevel-SingleExp_Illustration}.A), a similar value for the standard deviation (SD or $\sigma$) of the peak across channels is requested, ensuring comparable SO dynamics of the probed cell assemblies. Indeed, Down states are almost quiescent, and the variability of the MUA is mainly due to the acquisition chain, which has to be the same across channels.

A stable $\sigma$ allows the application of the same MUA threshold at each recording channel, meaning a unique definition of the Up states, and thus more reliable profiles of traveling wave-fronts and a more sound description of the SWA as a collective phenomenon. 
Therefore, the choice of a fixed \textit{UD\_THRESHOLD} is a valid option when the goal of the analysis is to study the dynamics of the propagation of the activity as a slow wave across the cortex. On the other end, this means to decide a key parameter \textit{a priori}, 
with the burden to be too conservative (increasing the false negative rate) or too tolerant (admitting a larger number of false positives)\footnote{In general, less conservative settings are preferable, since the control of false positives can be addressed by checking the spatio-temporal correlations of the propagating signal across the electrodes grid.}.
Conversely, the option of linking the choice of threshold to ``internal agents'' (\textit{e.g.} parameters evaluated during the workflow of the pipeline) can be convenient, for reducing the number of free variables, or for anchoring the false positive rate per channel.
A dedicated study of the standard deviation $\sigma$ has been carried out (Section \ref{subsec:SD-SummaryResults}) on the test-bench data.

Finally, channels not fulfilling the stability requirements are excluded from the analysis and marked as outliers. Here again, the decision on the stability requirements (which parameters to focus on, which threshold levels, which weight to assign at the different instances, how to define the outliers) is a key-element of the pipeline and may largely depend on specific features of the recording sessions and on the goals of the data analysis.
As discussed above, the SWAP pipeline has set in place alerts, warnings and counters based on parameters considered relevant for the test-bench data, but the elements to be monitored can change if conditions vary. 
Also configurable are the criteria that define outliers; for results presented here (Table \ref{tab:DAQsession-details}), excluded channels are those with the \textit{Right Peak} and with \textit{Few Transitions}, together with the requirement related to the stability of the Gaussian peak, that leaves out channels with $\sigma$ above $Q3 + 1.5 \times IQR$ ($IQR$ is inter-quartile range $Q3-Q1$, with $Q1$ first quartile and $Q3$ third quartile).
As indicated in the caption of Table \ref{tab:DAQsession-details}, the presence of discontinuity in the recording sequence -- corresponding to a failure in the data acquisition and a drop in the $log(MUA)$ -- is not a blocking element, since once the discontinuity is removed, the $log(MUA)$ can fulfil the stability requirements. 

\subsection{Stability of the data sample and channel selection/rejection}
\label{subsec:SD-SummaryResults}
\begin{figure}[h!]
    \centering
    \begin{overpic}[width=1\textwidth]{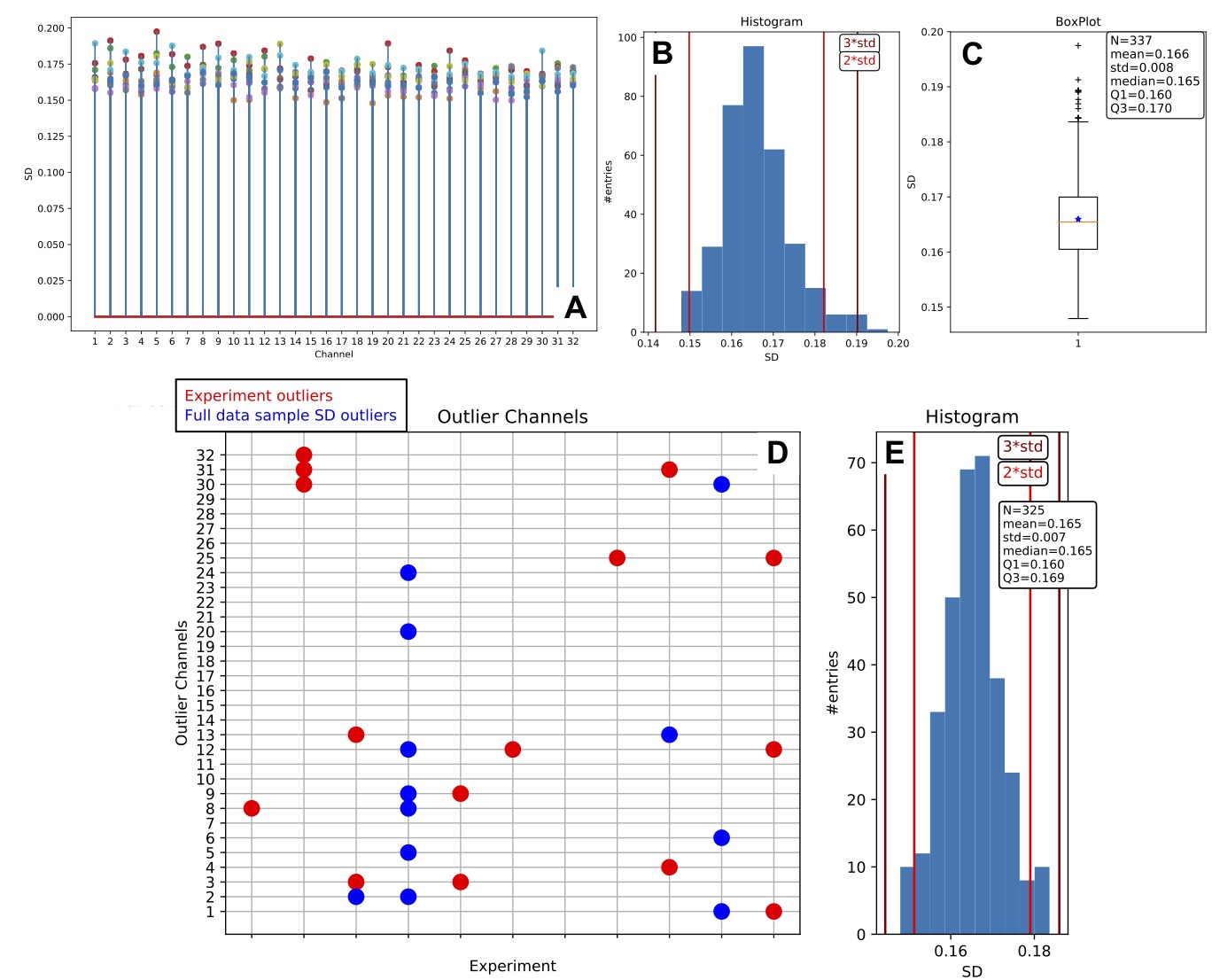}
        \put(20,2.2){\fontsize{8pt}{10pt}\selectfont{01\hspace{14pt}02\hspace{14pt}03\hspace{14pt}07\hspace{14pt}09\hspace{13pt}10\hspace{13pt}14\hspace{14pt}15\hspace{14pt}16\hspace{14pt}17\hspace{14pt}20}}
    \end{overpic}
\caption{Stability of the MUA estimate across channels and animals. \textbf{(A)} Stem plot of the SD values obtained as the $\sigma$ parameter of the Gaussian fit to the Down peak in the $log(MUA)$ distribution (the blue curve in Figure \ref{fig:ChannelLevel-SingleExp_Illustration}.A); marker colors identify the different experiments. The distribution of the SD values is presented as histogram \textbf{(B)} and as box plot \textbf{(C)}, the presence of outliers is evident; descriptive parameters are reported in the box; the vertical lines in the histograms indicate the positions displaced $2$ or $3$ standard deviations of the mean. Outlier values are removed, obtaining the symmetric distribution represented in the histogram in \textbf{(E)}; descriptive parameters are listed in the box, the sample of channels is reduced by $12$ units. The grid in \textbf{(D)} is a synthetical representation of the outlier channels for the test bench data: in red, the experiment outliers; in blue, the ones excluded after the SD Stability Study.}
\label{fig:SD-StabilityStudy}
\end{figure}

A key assumption for comparing acquisitions taken at different sites and with different electrodes (or from different animals) is the comparability of the $log(MUA)$ profile in the low firing rate regime. A quantitative evaluation of such a requirement is obtained by monitoring the width of the peak fitted with a Gaussian function, \textit{i.e.} the $\sigma$ parameter of the Gaussian (Figure \ref{fig:ChannelLevel-SingleExp_Illustration}.A). 
The purpose of this analysis step is to evaluate the selection strategy of the \textit{UD\_THRESHOLD}, which can be fixed or channel-dependent. 
A dedicated routine has been set in place, operating on a given collection of data (inter-session data from different subjects), aiming at stacking the full set of $\sigma$ parameters estimated from fitting procedures. 
In Figure \ref{fig:SD-StabilityStudy} we report the statistical distribution of $\sigma$ values, and we observe large stability across channels and experiments. 
In more detail, the plot offers an overview of the range of variability, showing that the behavior of the parameter is pretty stable inside each experiment, and stable in the entire collection as well\footnote{Experimental outliers, discussed in Section \ref{subsec:channel-level_SingleExp} and listed in Table \ref{tab:DAQsession-details}, are excluded from the representation.}.
Therefore, we adopted a channel-dependent \textit{UD\_THRESHOLD} set at $2\sigma$, corresponding to a fixed false positive rate per channel of about $2.25\%$. 
In addition, the results of the \textit{Stability Study} enable channel selection or rejection based on the entire data sample, giving rise to a further list of outliers, to be added to the one already filled out for each DAQ session at the end of the main loop. Considering both lists, a total of $27$ outliers were identified and removed from the analysis when inter-session data were taken into account for producing summary results.

\subsection{Description at the cortical area-level}
\label{subsec:area-level}

\subsubsection{A thumbnail for the single experiment}
\label{subsub:SigleExp_area-level}

Once the raw signals have been analyzed for each channel, area-related statistics can be obtained as in \citep{Ruiz-Mejias2011:AnesthetizedMouse}. More specifically, SWAP provides the following observables:
\begin{enumerate}
\item durations of the Down states, $d_{Down}[s]$;
\item durations of the Up states, $d_{Up}[s]$;
\item durations of the UD-cycles (\textit{i.e.} a pair of consecutive Down and Up states), $d_{UD}[s]$. The duration of the UD-cycle is an estimate of the SO period;
\item upward transition slope, $s_{Up}[s^{-1}]$;
\item downward transition slope $s_{Down}[s^{-1}]$;
\item maximum MUA in the Up state (``peak''), $p[a.u.]$
\item frequency, $f[Hz]$, defined as $\frac{1}{d_{UD}}$;
\end{enumerate} 

\begin{figure}[h!]
    \centering
    \begin{overpic}[width=1\textwidth,trim={0.25 -0.1cm 0.25 0cm},clip]{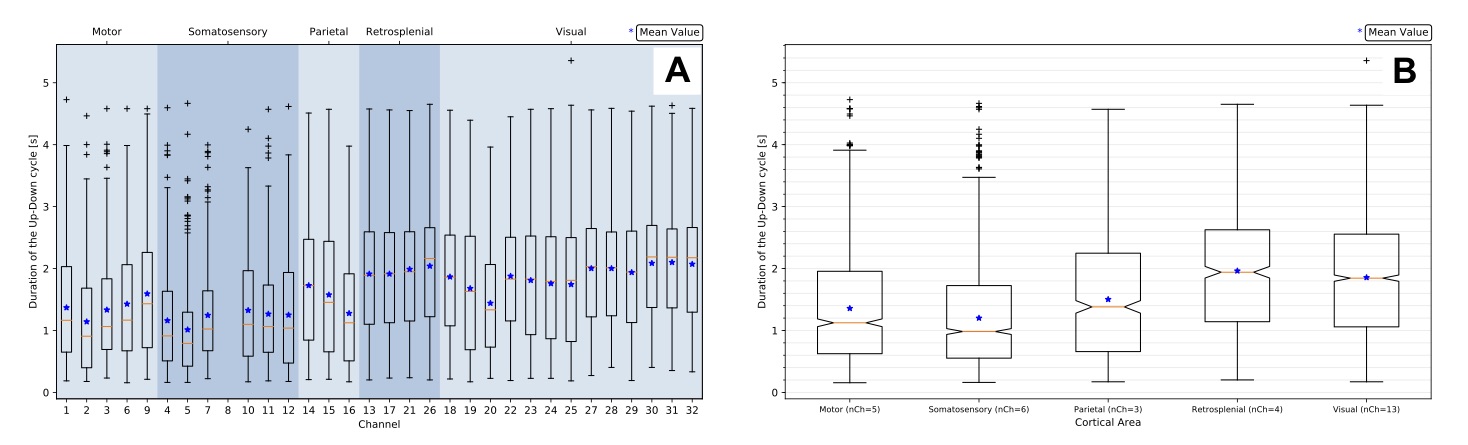}
        \put(55.5,1.6){\fcolorbox{white}{white}{\fontsize{5pt}{7pt}\selectfont{Motor\hspace{18pt}Somatosensory\hspace{12pt}Parietal \hspace{16pt}Retrosplenial\hspace{16pt}Visual}}}
        \put(56,0){{\fontsize{5pt}{7pt}\selectfont{(nCh=5)\hspace{21pt}(nCh=6)\hspace{20pt}(nCh=3)\hspace{22pt}(nCh=4)\hspace{21pt}(nCh=13)}}}
        \put(5.4,28.5){\fcolorbox{white}{white}{\fontsize{6pt}{8pt}\selectfont{M1\hspace{30pt}S1\hspace{22pt}PtA \hspace{10pt}RSC\hspace{42pt}V1}}}
    \end{overpic}
\caption{Statistics of the measured UD-cycle ($d_{UD}$) for an example recording (file 01).  
\textbf{(A)} Channel-level description; channels are grouped by area, channel numbering is as illustrated in Figure\ref{fig:ElectrodeArray}; channel 8 is missing because it is tagged as an outlier (Table \ref{tab:DAQsession-details}). \textbf{(B)} Area-level description, represented with notches indicating the confidence interval for the median ($Q2 = 50\%$, orange line). The box plot is delimited by quartiles $Q1 = 25\%$ and $Q3 = 75\%$; $IQR = Q3-Q1$ is the \textit{Inter Quartile Range}; the lower whisker is at $Q1-1.5 \times IQR$; the upper whisker is at $Q3+1.5 \times IQR$; values outside the whiskers are marked as outliers; the blue marker indicates the mean of the distribution.}
\label{fig:DurationSingleExpBoxPlot}
\end{figure}

Concerning $d_{Down}$, $d_{Up}$ and $d_{UD}$, as already evident from the histograms (Figure \ref{fig:ChannelLevel-SingleExp_Illustration}.C) and in agreement with the box plots represented in Figure \ref{fig:DurationSingleExpBoxPlot}.A, the distributions at the channel level are skewed and far from being Gaussian. Therefore, for each channel, the median (and not the mean) has been selected as the representative parameter for the observables. The skewness of the distribution is also noticeable at the area level (Figure \ref{fig:DurationSingleExpBoxPlot}.B), where the statistics for each distribution are increased since values of electrodes belonging to the same cortical area are grouped together. Therefore, the median is assumed as the representative parameter also at the area level.

\begin{figure}[h!]
    \centering
    \begin{overpic}[width=0.7\textwidth,trim={0 0cm 0 -2cm},clip]{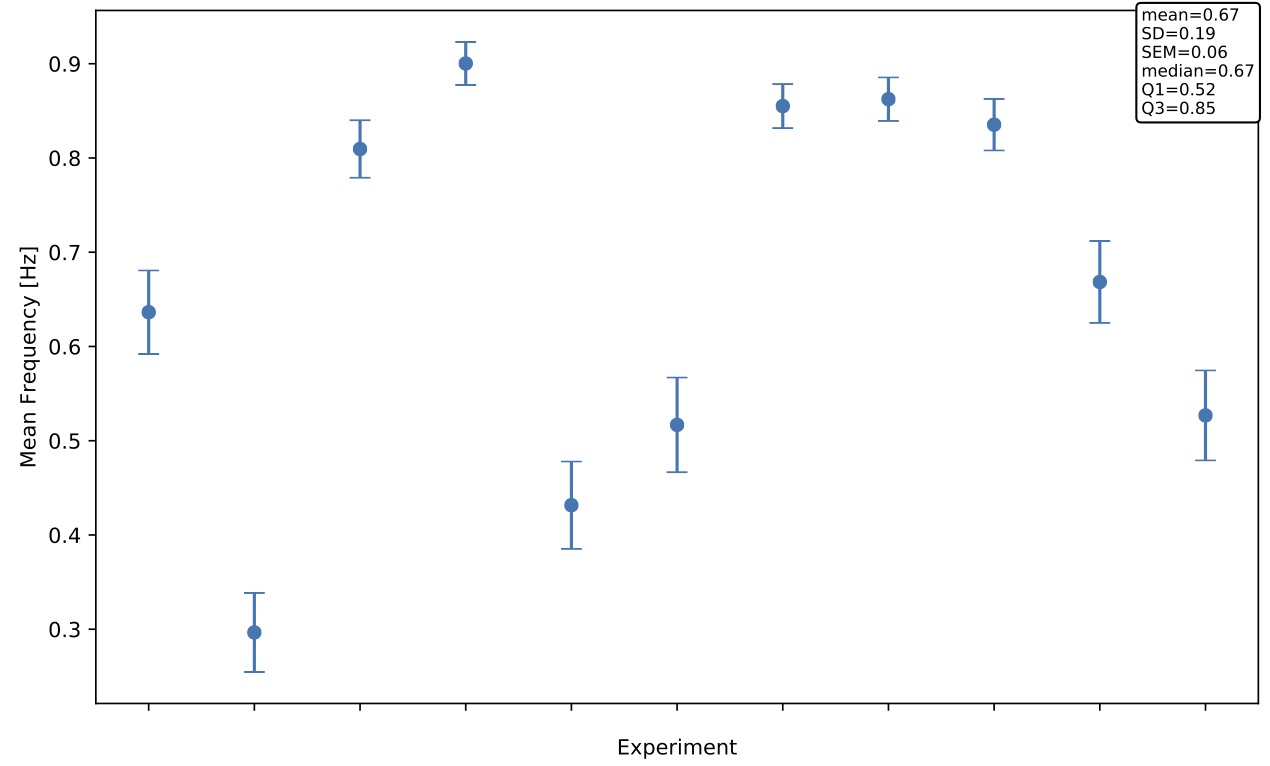}
        \put(11,2.5){\fontsize{6pt}{8pt}\selectfont{01\hspace{20pt}02\hspace{20pt}03\hspace{20pt}07\hspace{20pt}09\hspace{19pt}10\hspace{19pt}14\hspace{20pt}15\hspace{20pt}16\hspace{20pt}17\hspace{20pt}20}}
    \end{overpic}
\caption{Mean frequency across channels for each experiment; error bars, SEM (standard error of the mean). For each channel, the frequency $f[Hz]$ is computed as $\frac{1}{<d_{UD}>}$, with $<>$ denoting the arithmetic mean of the $d_{UD}[s]$ values in the channel ($<d_{UD}[s]>$ is an estimate of the SO period). The text box gives information on the distribution of the observable across the $11$ experiments.} 
\label{fig:MeanFrequencyPlot}
\end{figure}

Interestingly, significant differences are apparent when comparing median values of different channels and areas. 
Differences among areas are consistently observed in all animals, despite the large span of values that each observable expresses considering the total of 11 experiments.
Indeed, evaluating each observable for each experiment at the full-set level of description, the large variability of the data sample is clearly seen. This is in agreement with the expected biological variability of the subjects, regardless of the identical surgical treatment they have been undergone,  the comparable drug delivery, and the uniform monitoring conditions during the data taking. In other words, each data session is characterized by its own central values for all the observables of interest, with no clear correlation with the animal's phenotype, or with any other parameter measurable during the data acquisition \citep{Brown2010:AnesthesiaSleepComa,Brown2017:Anesthesia-dsitinct-from-Sleep}. The spanning of the frequency values across the experiments (Figure \ref{fig:MeanFrequencyPlot}) is in particular revealing, because frequency is a property immediately linkable with the onset of the bimodality and with the propagation of slow waves along the cerebral cortex.

\subsubsection{Assessment of statistical significance at the area-level description for inter-session normalized data}
\label{subsub:assessment_area-level}
\begin{figure}[h!]
    \centering
    \begin{overpic}[width=0.9\textwidth,trim={0cm 0cm 18cm 0cm},clip]{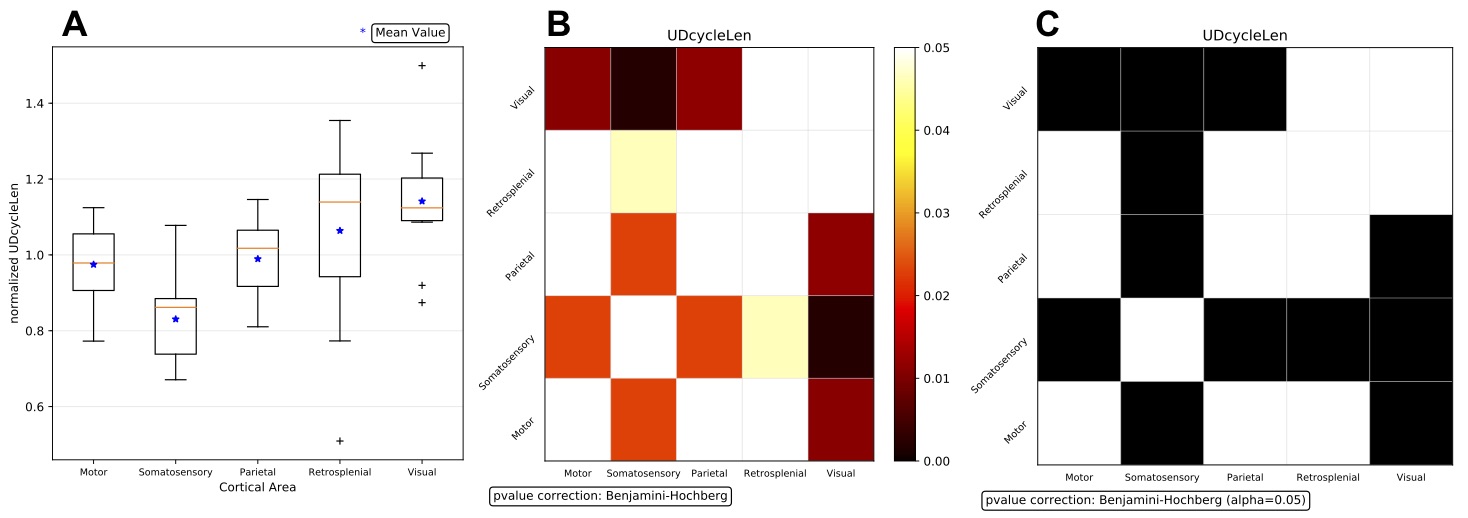}
        \put(6,51){\fcolorbox{white}{white}{\fontsize{11pt}{13pt}\fontfamily{qhv}\selectfont{\textbf{A}}}}
        \put(57,51){\fcolorbox{white}{white}{\fontsize{11pt}{13pt}\fontfamily{qhv}\selectfont{\textbf{B}}}}
    \end{overpic}
\caption{\textbf{(A)} Box plot representation of the summary results for the observable $d_{UD}$ at the area-level description, obtained with normalized data. Having $11$ subjects, for each cortical area $11$ median values are available (the orange line in Figure \ref{fig:DurationSingleExpBoxPlot}); the arithmetic mean of the $5$ medians is the normalization factor for the given experiment. 
\textbf{(B)} Outcome of the Wilcoxon hypothesis tests on the same data; the test results are represented via a matrix of \textit{p}-values (with Benjamini-Hochberg correction enabled).}
\label{fig:DurationAreaLevelSummaryResultsPvalMatrix}
\end{figure}

The statistical significance of the effect of differentiation by area, visible for all the observables in the list of interest and for each experiment in the data set, is quantitatively assessed for summary results, \textit{i.e.} when the outcome of the single experiments are combined to drive comprehensive claims on the phenomena. 
However, the first obstacle met when trying to compare and aggregate results is the large variability exhibited by the different DAQ sessions, which dominates over any other effect and disguises any similarity or common footprint among subjects.   
Therefore, to confront the area-level descriptions, the median values of the observables for each area are normalized for each experiment, computing for each observable the arithmetic mean across the cortical areas, and using the obtained mean as a normalization factor for that observable. This procedure highlights any trend expressed at different cortical areas, enabling us to check if different experiments express the same trend. 
Figure \ref{fig:DurationAreaLevelSummaryResultsPvalMatrix}.A shows the summary result at the area-level description obtained with normalized data.  

The outcome of the hypothesis tests executed to assess the statistical significance of the differences observed between each couple of cortical areas is represented through a matrix of \textit{p}-values, with a graded scale in a given range of confidence (Figure \ref{fig:DurationAreaLevelSummaryResultsPvalMatrix}.B).
Since no assumptions are made on the model, and because of the already-discussed evidence of non-Gaussianity of the distributions, non-parametric approaches are followed when analyzing the test bench data, so the Wilcoxon rank-sum test is applied\footnote{The computation of the Wilcoxon rank-sum statistics is carried out with the statistical function \texttt{scipy.stats.ranksums} of the \texttt{scipy} Python module \cite{Python:SciPy}}. 
To take into account multiple comparisons (the so-called ``look-elsewhere effect''\footnote{\url{https://xkcd.com/882/}.}) and reduce the likelihood of incorrectly rejecting a null hypothesis (type I error) when evaluating a family of simultaneous tests, the robustness of the analysis and the control of the false discovery rate (FDR) is obtained by correcting the \textit{p}-values with the Benjamini-Hochberg (BH) procedure \citep{Seabold2010:statsmodels}.

Analogously to what was done for state duration, the study of a possible differentiation at the area-level has also been carried out for slopes and maximum MUA ($s_{Up}$, $s_{Down}$, $p$).
Slopes are computed considering the average transition (Figure \ref{fig:ChannelLevel-SingleExp_Illustration}.F--G), obtained pooling together the detected Up-to-Down and Down-to-Up transitions. 
The transition front of the average transition is isolated, considering a $35\ ms$ interval around the transition time $t_0$ ($[t_0-0.025, t_0+0.010]$ for downward transitions; $[t_0-0.010, t_0+0.025]$ for upward transitions); the profile is fitted with a cubic and the slope is obtained as the derivative of the polynomial at $t_0$. The average upward transition is also used for the estimation of the maximum MUA of the average waveform in the first $250\ ms$ after the transition.
Since slopes and maxima are calculated from the average waveform, for each experiment the observables are represented at the channel level by a single value (and not by a distribution). 
The area-level description for the single experiment is given by the mean and median of values obtained from all the channels belonging to the specific area.
In coherence with the analysis carried out on states duration and in order to apply the same non-parametric Wilcoxon tests, the median is taken as the representative parameter. Summary results are produced after the same normalization adopted for states duration, yielding a similar illustration.
A similar approach is followed for frequency $f$. 

\subsection{Interpolation of the array map}
\label{subsec:arraymap}

\begin{figure}[h!]
    \centering
    \begin{overpic}[width=0.9\textwidth]{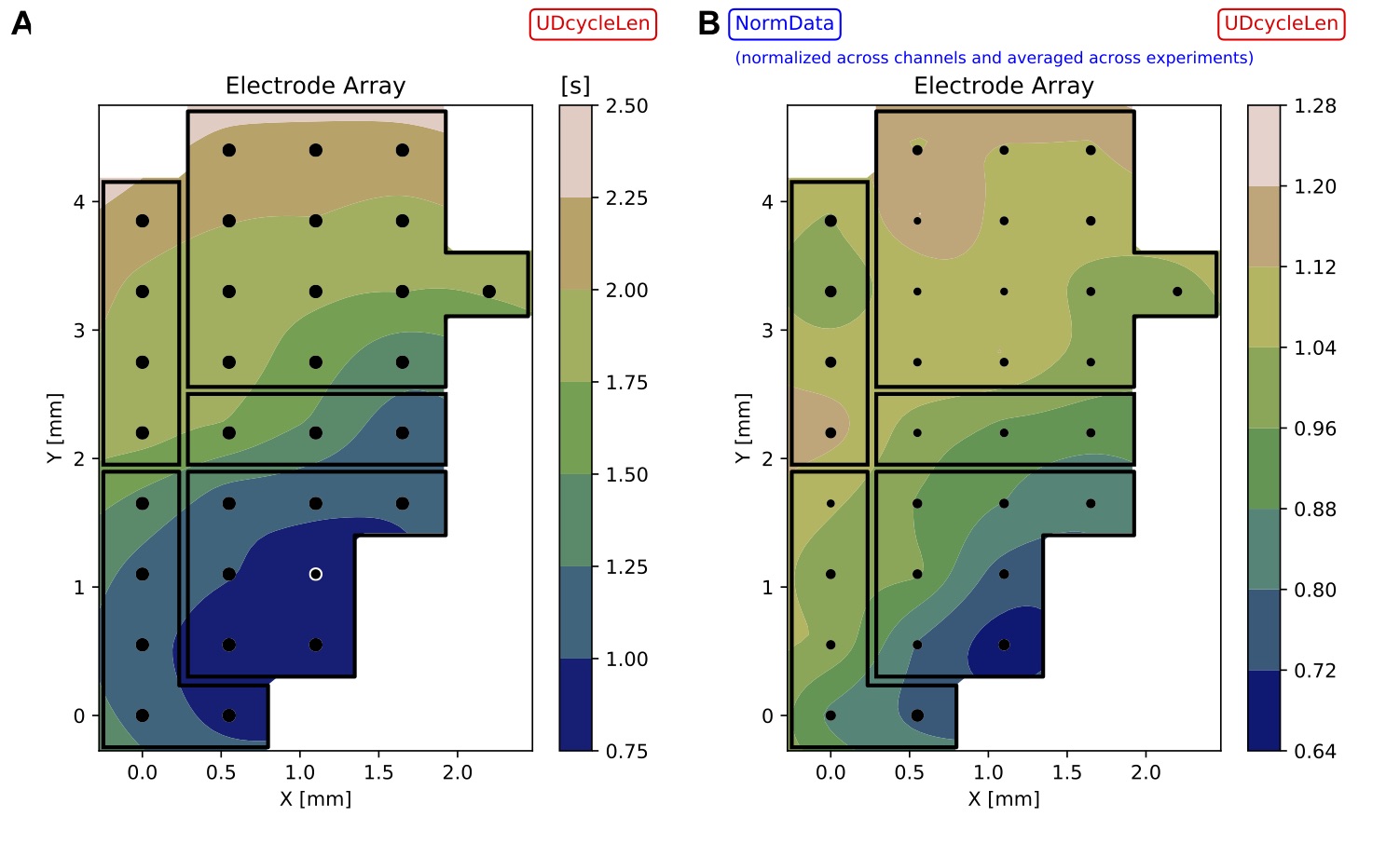}
        \put(16,0){\fontsize{7pt}{9pt}\selectfont{medial $\longleftrightarrow$ lateral}}
        \put(0,21){\rotatebox{90}{\fontsize{7pt}{9pt}\selectfont{anterior $\longleftrightarrow$ posterior}}}
        \put(-1,57){\fcolorbox{white}{white}{\fontsize{11pt}{13pt}\fontfamily{qhv}\selectfont{\textbf{A}}}}
    \end{overpic}
\caption{\textbf{(A)} Contour plot for the single experiment (here, file 01) illustrating the observable $d_{UD}$ (\textit{UDcycleLen}) interpolated on the mesh-grid, based on values recorded with the MEA. The array map reflects the DAQ schema of Figure \ref{fig:ElectrodeArray}, with black dots marking the electrode positions and black lines identifying cortical areas; the white circle identifies the outlier channel as resulting from Figure \ref{fig:SD-StabilityStudy} and Table \ref{tab:DAQsession-details}. The color bar gives the range of the interpolated variable across the array. \textbf{(B)} Summary results for the observable $d_{UD}$, represented as a contour plot obtained from normalized data. The interpolation is based on the mean values computed across the experiments, the marker size is inversely proportional to the standard deviation of the distribution of values at each position (the smaller the marker, the less variability measured across the experiments). Since normalized data are used for the plot, the color bar gives indications of trends with respect to the average: for instance, regions colored in blue are characterized by having a duration of the oscillation cycle that is up to $30\%$ less than the average. Comparing \textbf{(A)} (single experiment) and \textbf{(B)} (inter-session data), a similar tendency is visible for the observable: in particular, the contours evidence gradients along the same dominant antero-posterior direction, from fronto-lateral towards occipito-medial regions. These gradients prevail over the borders of cortical areas, suggesting further differentiation within areas and connections among areas.}
\label{fig:UD_SingleExp-and-SummaryRes_Contour}
\end{figure}

The results obtained with the high spatial resolution probe used for the acquisition of the test bench data can still be improved by interpolating the spaces between the electrodes; the accuracy of the interpolation is assured by the large surface density of sensors offered by the employed MEA. 
For each experiment and channel, the median ($d_{Down}$, $d_{Up}$, $d_{UD}$) or the mean ($s_{Up}$, $s_{Down}$, $p$, $f$) of the observable is taken. 
This set of values is used to compute an interpolator\footnote{The interpolation is carried out with the \texttt{scipy} Python module \cite{Python:SciPy} using the \texttt{scipy.interpolate.Rbf} class for radial basis function (Rbf) interpolation. 
More on \url{https://docs.scipy.org/doc/scipy/reference/generated/scipy.interpolate.Rbf.html}} 
that is applied to points ($xy$ coordinates) on a mesh-grid, made of 50 steps along $x$ and 90 step along $y$, with a \textit{Grid Step} $\sim 0.05 \: mm$, \textit{i.e.} $1/10$ of the \textit{Array Step} $=0.550 \: mm$ (Figure \ref{fig:UD_SingleExp-and-SummaryRes_Contour}.A for the single experiment and for a given observable).
The same illustration is applied when representing inter-session normalized data (Figure \ref{fig:UD_SingleExp-and-SummaryRes_Contour}.B); here, the size of the marker is inversely proportional to the standard deviation of the distribution of values at the given electrode position, thus measuring the amount of variability registered at that position across the experiments. 

\subsection{From the array map to the assessment of statistical significance at the channel level for the normalized inter-session data}
\label{subsec:Stats_channel-level}

The contour plots traced on the interpolated array maps give qualitative hints on differentiation among inspected cortical sites, enlightening further details within cortical areas.
A more quantitative evaluation can be obtained by performing the statistical analysis of the normalized values resulting from inter-session data, considering each pair of electrodes in the MEA and inspecting if there is a statistically significant difference between the distribution of values measured therein.
The idea is to compare the normalized values located at two different electrode positions in the map and to test the validity of the null hypothesis, that is the samples are extracted from the same statistical distribution\footnote{The number of samples at each location depends on how many experiments have channels rejected as outliers at that location. For the test bench data set, a maximum of $11$ samples can be present at each electrode position, details for each electrode position can be extracted from Figure \ref{fig:SD-StabilityStudy}.D.}. We use the Wilcoxon rank-sum statistics with B-H correction on \textit{p}-values; in this case, the FDR correction has a larger impact, since for MEAs the number of tests in the family is of the order of hundreds\footnote{If the number of electrodes in the MEA is $n=32$, the number of hypotheses to be checked is $N=\frac{n(n-1)}{2}=496$.}.

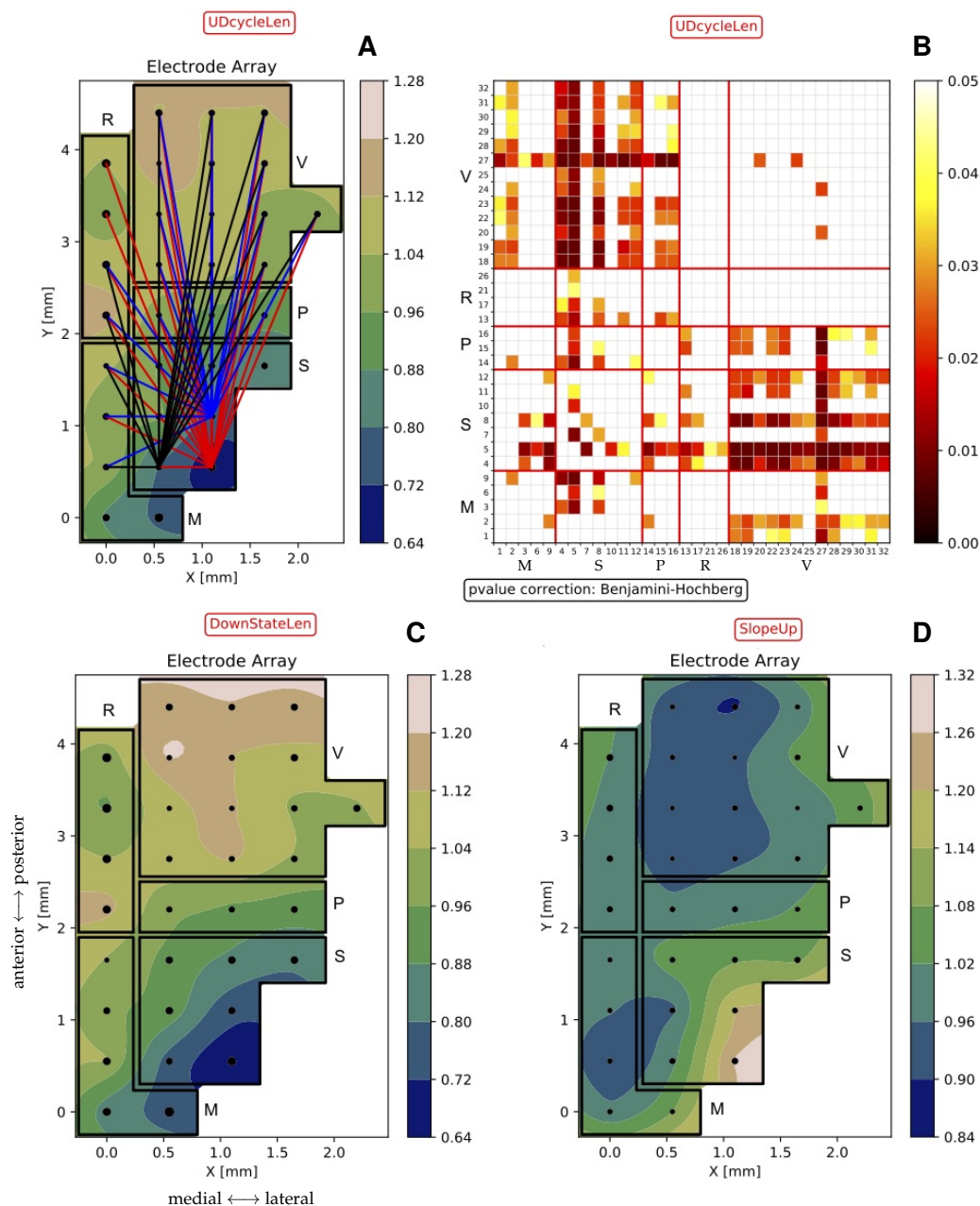
\begin{figure}[h!]
    \centering
    \begin{overpic}[width=0.9\textwidth]{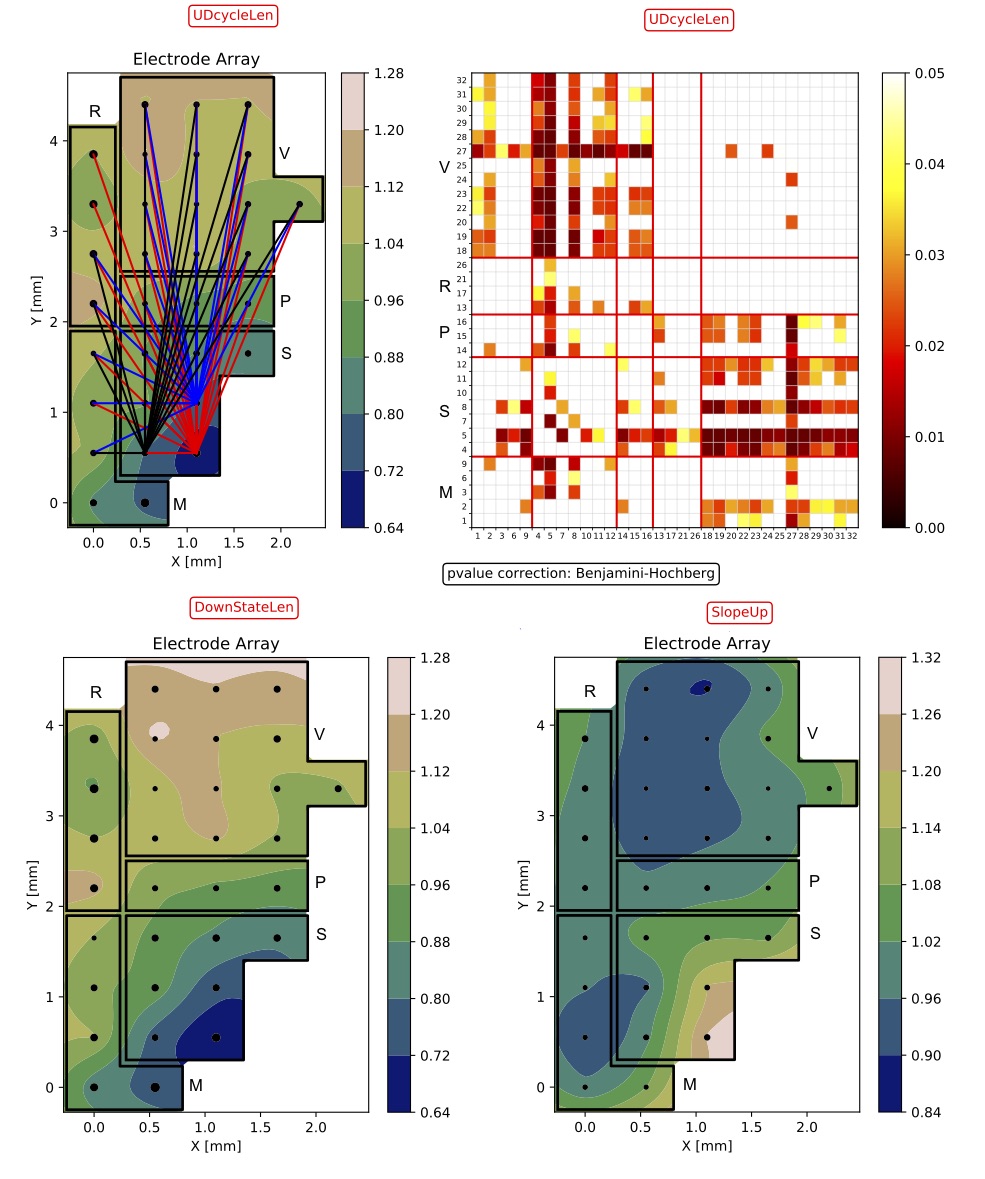}
        \put(42,53){\fontsize{6pt}{8pt}\selectfont{M\hspace{25pt}S\hspace{21pt}P\hspace{14pt}R\hspace{36pt}V}}
        \put(13,0){\fontsize{7pt}{9pt}\selectfont{medial $\longleftrightarrow$ lateral}}
        \put(0,18){\rotatebox{90}{\fontsize{7pt}{9pt}\selectfont{anterior $\longleftrightarrow$ posterior}}}
        \put(28,96){\fcolorbox{white}{white}{\fontsize{11pt}{13pt}\fontfamily{qhv}\selectfont{\textbf{A}}}}
        \put(74,96){\fcolorbox{white}{white}{\fontsize{11pt}{13pt}\fontfamily{qhv}\selectfont{\textbf{B}}}}
        \put(32,47){\fcolorbox{white}{white}{\fontsize{11pt}{13pt}\fontfamily{qhv}\selectfont{\textbf{C}}}}
        \put(74,47){\fcolorbox{white}{white}{\fontsize{11pt}{13pt}\fontfamily{qhv}\selectfont{\textbf{D}}}}
    \end{overpic}
\caption{Excitability maps of the probed cortices. \textbf{(A)} Normalized $d_{UD}$ (as in Figure \ref{fig:UD_SingleExp-and-SummaryRes_Contour}.B) with the ``core-nodes'' (\textit{i.e} the three electrodes with the largest number of significant differences with the other channels). Lines are traced to connect the ``core nodes'' with their partners in the electrode pair being significantly different ($p < 0.05$). \textbf{(B)} Matrix of \textit{p}-values of the statistical test on the $d_{UD}$ differences between channels. The matrix construction is analogous to what is described for Figure \ref{fig:DurationAreaLevelSummaryResultsPvalMatrix}. The numeric sequence of electrodes in the matrix is as in Figure \ref{fig:DurationSingleExpBoxPlot}, \textit{i.e.} channels are grouped by area (M = motor, S = somatosensory, P = parietal, R = retrosplenial, V = visual). \textbf{(C,D)} Normalized Down state duration $d_{Down}$ (\textit{DownStateLen}) and Down-to-Up transition slope $s_{Up}$ (\textit{SlopeUp}), averaged across experiments, as in Panel \textbf{(A)}.}
\label{fig:UD_SummaryRes_Contour_pvalMatrix}
\end{figure}

Given the large number of hypotheses, electrodes are sorted to emphasize those expressing a significant difference with the others. More specifically, the top-three electrodes in this sorted list are highlighted as ``core nodes'', indicating the positions in the cortex displaying the largest significant difference with respect to other positions in the cortex  (Figure \ref{fig:UD_SummaryRes_Contour_pvalMatrix}.A).

As a result, the somatosensory cortex emerges as the one having the shortest mean Up/Down cycle, in agreement with what shown in Figure \ref{fig:DurationAreaLevelSummaryResultsPvalMatrix}.A. In Figure \ref{fig:UD_SummaryRes_Contour_pvalMatrix}.B, the differences in the cycle duration at the channel level show a rather homogeneous rhythm within single areas. It is also apparent that the other primary  sensory area, \textit{i.e.} the visual cortex, is markedly different from frontal and parietal regions, displaying a significantly longer oscillation period. Inspecting the changes in the Up and Down state duration, we found that the modulation of the oscillation cycle is mainly due to a reduction of the $d_{Down}$, as shown in Figure \ref{fig:UD_SummaryRes_Contour_pvalMatrix}.C. Indeed, only few channels display a significant difference in $d_{Up}$ (not shown) and the p-value matrix for the $d_{Down}$ mirrors the one shown in Figure \ref{fig:UD_SummaryRes_Contour_pvalMatrix}.B. 
If the local excitability of the cell assemblies underneath the MEA contacts is well represented by the ratio $d_{Up}/d_{Down}$ \citep{Ruiz-Mejias2011:AnesthetizedMouse,Mattia2012:ExploringTheSpectrum}, here the leading role of somatosensory cortex is apparent.
Such enhanced excitability is further confirmed by inspecting the slope of the Down-to-Up transition in the MUA \citep{SanchezVives2010:InhibitoryModulation}, which is significantly steeper in the same cortical locations (Figure \ref{fig:UD_SummaryRes_Contour_pvalMatrix}.D).

\section{Discussion}
\label{sec:discussion}

Biological data are characterized by richness in details and large variability. The efforts of the data analysis should aim at extracting tendencies and regularities, producing a concise description without hiding or neglecting complexity and details that could convey informative content. This is the guideline followed when developing the Slow Waves Analysis Pipeline (SWAP), starting from a solid backbone that has been deeply revised and enriched with new features. In particular, the opportunity of using the MEA data as a test bench has allowed us to focus on the spatial differentiation of the observables, with the aim of uncovering hints as to the local excitability of the cortical assemblies.
The developed methodology is robust and easily re-configurable, flexible and adaptable at different acquisition conditions, and also suitable to be applied to the output of simulations. 
In this framework, SWAP can be employed in bridge theory, simulations and experiments, providing a set of general tools that allow an effective comparison between heterogeneous data sets.
The adoption of a unique analysis procedure is also useful for comparing different simulation engines; the SWAP can be applied to define benchmarks and evaluate the performance of numerical models and implementations.
Several studies are ongoing for the application of SWAP to a large variety of datasets (knock-out mice, subjects in different brain states, data collected with optical techniques), and the stability and reliability of the analysis procedure has been so far confirmed.
The introduced new features in the analysis pipeline have been coded in Python with the aim of realizing open software tools for the scientific community. The complete transition towards open software is in the list of objectives to fulfill in the near future.

Concerning the interpretation of the results, the analysis of the large set of data collected from 11 wild-type anesthetized mice with the 32-channel MEA allows us to claim a statistically significant differentiation of cortical areas for several parameters that characterize the onset of SWA along the cerebral cortex. 
Starting from these observables, the excitability of the cortical tissue expressing SWA can be investigated. Indeed, larger excitability is expected to be associated with faster transition fronts (in particular, upward slopes), shorter duty-cycles (\textit{i.e.} smaller $d_{UD}$, dominated by the duration of the Down states) and accordingly larger frequencies \citep{SanchezVives2010:InhibitoryModulation}. 
For instance, a smaller $d_{Down}$ reveals the case in which excitability translates in faster Down-to-Up transitions. These features are particularly apparent in the somatosensory area, likely  the most excitable cortical region we observed; conversely, the occipital lobe (retrosplenial and visual areas) act as the least excitable.
Activation waves traveling across the cortex during SWA are expected to be sensitive to the diverse cortical excitability, as more reactive (\textit{i.e.} more excitable) areas are expected to display a smaller ``inertia'' in recruiting neurons involved in the collective phenomena associated with the high-firing Up states. As a consequence, waves might be originating from highly excitable regions such that preferential propagation pathways are expected, as previously highlighted both in humans \citep{Massimini2004:SleepTravelingWave} and rodents \citep{Ruiz-Mejias2011:AnesthetizedMouse,Stroh2013:MakingWaves,Mohajerani2013:SpontaneousCorticalActivity}. In turn, heterogeneous excitability might also underlie the non-global nature of the SWA phenomenon observed when wakefulness is approaching \citep{Vyazovskiy2011:LocalSleep,Nir2011:RegionalSlowWaves} or under pathological conditions \citep{SanchezVives2017:DefaultActivityPattern}.

It must be pointed out that the study here presented is ``static'', in the sense that the absolute time sequence of the events is not taken into account: states and transitions are time-squashed and stacked regardless of their time of occurrence in the DAQ session, and thus any time-dependent effect, like fatigue and recovery intervals of the neurons, that could affect the excitability and alter the responsiveness of cortical regions, is excluded from this analysis. In other words, the figures presented in this paper can be seen as the average photography of the phenomena investing the monitored cortex during the acquisition period. 
Related to this, excitability and responsiveness can also be altered by wave propagation dynamics, in particular if different propagation patterns coexist and overlap in the same time interval, for instance when two slow waves originating at distinct sites travel along the cortex at different speeds and each one follows its own path \citep{Sancristobal2016:CollectiveStochasticCoherence,Capone2017:SlowWavesCorticalSlices}.
Again, not using the information on the absolute timing of the events at the electrode positions and intending the results presented here to be the average photography of the SWA in the cortex, the included results reflect the dominant propagation pattern, namely the antero-posterior direction \cite{Massimini2004:SleepTravelingWave,Stroh2013:MakingWaves,Ruiz-Mejias2011:AnesthetizedMouse,Chan2015:MesoscaleInfraslow}, in particular along an oblique axis directed from fronto-lateral towards occipito-medial regions, as suggested by values depicted in the contour plots.
As in this step of the SWAP we have not focused on time-dependent effects, their impact on the area-differentiation of the bistability modes will be evaluated in further studies. 

Finally, although in all the animals involved in this study the level of administered anesthesia was the same, the observed inter-subject variability (Figure \ref{fig:MeanFrequencyPlot}) could in principle be explained by animals being in different brain states. This suggests the possibility of exploiting the characterization of slow oscillations in the cerebral cortex as a new tool for effective classification of the brain states.
Several techniques are currently under study, based for instance on the principal components analysis (PCA) to identify and single out the different sources of variability in the experimental data set. 
Indeed, a more reliable classification of brain states (\textit{i.e.} of the DAQ sessions that constitute the statistical sample) would provide a more robust comparison of the experiments, allowing us to overcome the limits derived from the need to use normalized data.

\section*{Conflict of Interest Statement}
The authors declare that the research was conducted in the absence of any commercial or financial relationships that could be construed as a potential conflict of interest.
\section*{Author Contributions}
MVSV, MM, MD, PSP and GDB  conceived and designed the study. 
MD performed the data acquisition. 
MM provided the backbone of the analysis pipeline.
GDB revised the analysis pipeline, improved it with new functions, applied it to the test bench data, performed the statistical analysis, wrote the first draft of the manuscript. 
PSP and MM contributed to the data analysis and to the interpretation of the results.
AP revised sections of the manuscript.
All authors contributed to manuscript revision, and read and approved the submitted version.
\section*{Funding}
This work has received funding from the European Union Horizon 2020 Research and Innovation Programme under Specific Grant Agreements No. 785907 (HBP SGA2) and No. 720270 (HBP SGA1).
\section*{Acknowledgments}
This study was carried out in the framework of the Human Brain Project (HBP\footnote{\url{https://www.humanbrainproject.eu/en/}}), funded under Specific Grant Agreements No. 785907 (HBP SGA2) and No. 720270 (HBP SGA1), in particular within activities of sub-project SP3 (``Systems and Cognitive Neuroscience'').
\section*{Data Availability Statement}
Data will be available on reasonable request.




\bibliographystyle{frontiersinHLTH_FPHY} 
\bibliography{myBibHBP}



\end{document}